# Compressive strain-induced metal-insulator transition in orthorhombic SrIrO$_3$ thin films

J. H. Gruenewald, J. Nichols, J. Terzic, G. Cao, J. W. Brill, and S. S. A. Seo

*Department of Physics and Astronomy, University of Kentucky, Lexington, KY 40506*

We have investigated the electronic properties of epitaxial orthorhombic SrIrO$_3$ thin-films under compressive strain. The metastable, orthorhombic SrIrO$_3$ thin-films are synthesized on various substrates using an epi-stabilization technique. We have observed that as in-plane lattice compression is increased, the *dc*-resistivity ($\rho$) of the thin films increases by a few orders of magnitude, and the d$\rho$/d$T$ changes from positive to negative values. However, optical absorption spectra show Drude-like, metallic responses without an optical gap opening for all compressively-strained thin films. Transport measurements under magnetic fields show negative magneto-resistance at low temperature for compressively-strained thin-films. Our results suggest that weak localization is responsible for the strain-induced metal-insulator transition for the orthorhombic SrIrO$_3$ thin-films.

## I. INTRODUCTION

The 5$d$ transition metal oxides have drawn much interest for their exotic phases arising from the interplay between strong spin-orbit interaction and electronic correlation.[1-3] These materials, which include the iridates, were originally predicted to be weakly correlated, paramagnetic metals due to the extended nature of the 5$d$ orbitals and a partially-filled valence band. However, a few interesting phases have been theoretically proposed for systems exhibiting both strong spin-orbit coupling and strong electron correlation. The iridates are prime candidates for realizing these unusual states, such as topological insulators, unconventional superconductivity, quantum spin-Hall effect, and Weyl semimetals to name a few.[4-7] One system of recent interest is the Ruddlesden-Popper (R-P) series iridates, $Sr_{n+1}Ir_nO_{3n+1}$, whose electronic structure is tunable from a three-dimensional correlated metal, $SrIrO_3$ ($n = \infty$), to a two dimensional $J_{eff} = 1/2$ Mott insulator, $Sr_2IrO_4$ ($n = 1$).[1] The insulating state emerges when an octahedral crystal field splits the degenerate 5$d$ levels into $e_g$ and $t_{2g}$ bands; the partially filled $t_{2g}$ bands ($L_{eff} = 1$) are further split into the $J_{eff} = 3/2$ and $J_{eff} = 1/2$ bands by the strong spin-orbit coupling inherent in iridium ions. The Mott gap opens in the $J_{eff} = 1/2$ band if the on-site Coulomb interaction becomes energetically comparable to or larger than the bandwidth. In $SrIrO_3$, there are six nearest neighbor Ir atoms, while in $Sr_2IrO_4$ there are only four. This reduction of coordination number of Ir 5$d$ orbitals reduces the bandwidth and opens a gap in the partially filled $J_{eff} = 1/2$ band in $Sr_2IrO_4$. Hence, the metal-insulator transition in the R-P series iridates is driven by a *dimensionally controlled decrease in bandwidth*.

In this paper, we have investigated whether a metal-insulator transition can also occur in epitaxial $SrIrO_3$ thin films via a strain-induced reduction in bandwidth. While the Ir-O bond length is rigid and difficult to change, the Ir-O-Ir bond angle can be readily affected by lattice strain.[8,9] As schematically illustrated in Fig. 1, when the Ir-O-Ir bond angle is decreased by compressive strain, the electronic hopping integral between Ir $5d$ orbitals is reduced, and hence the bandwidth decreases. This reduction in bandwidth can induce a metal-insulator transition either by opening a Mott gap, similar to $Sr_2IrO_4$,[1] or by creating a localized band below the mobility edge due to disorder.[10] We have grown epitaxial $SrIrO_3$ (SIO) thin films on various substrates with close lattice mismatch. We find that as in-plane compression increases, the metallic SIO thin films become insulating. Optical absorption spectroscopy indicates that all SIO thin films have a metallic Drude-like response and lack the presence of an optical gap, despite the insulating transport behavior. Transport measurements under magnetic field display a negative magneto-resistance (MR) behavior in the insulating SIO thin films. Hence, the strain-induced metal-insulator transition is due to weak localization. Our results demonstrate that the electronic properties of perovskite SIO thin films can be manipulated by compressive lattice strain.

## II. SAMPLE SYNTHESIS

We have grown orthorhombic SIO epitaxial thin films on several substrates with various compressive strains using pulsed laser deposition. The thermodynamically stable phase of bulk SIO crystals is the hexagonal $6H$-$BaTiO_3$ structure at ambient conditions, and the orthorhombic phase can be stabilized using a high-pressure synthesis.[11,12] However, orthorhombic SIO phases can be stabilized as epitaxial thin films on pseudo-cubic perovskite substrates.[13,14] Four single-crystal substrates are used: MgO (001), $GdScO_3$ (110) (GSO), $SrTiO_3$ (001) (STO), and $(LaAlO_3)_{0.3}(Sr_2AlTaO_6)_{0.7}$ (001) (LSAT) with in-plane pseudo-cubic lattice constants of 4.21, 3.96, 3.90, and 3.87 Å, respectively. During the deposition, the substrate temperature is 700 °C in an oxygen partial pressure of 100 mTorr. The samples are post-annealed and cooled in a 1 Torr oxygen atmosphere.

Figure 2a shows $\theta$-$2\theta$ X-ray diffraction data for all the SIO thin films used in this study, and all thin films have the correct *00l* orientation with respect to the substrate normal. As the in-plane lattice parameters are decreased from MgO to LSAT, the *00l* thin film peaks are shifted to lower angles, indicating an out-of-plane (*c*-axis) expansion of the SIO unit cell. This is consistent with an in-plane compressive strain causing an elongation along the *c*-axis. The SIO thin film thicknesses are found to be around 20 nm using X-ray reflection fringes. The rocking curves in Fig. 2b are taken of the *002* film reflection. The thin films grown on GSO, STO, and LSAT have full widths at half maximum less than 0.08°, indicating high crystallinity. X-ray reciprocal space maps of the (pseudo-cubic) 103-reflection in Fig. 2c show the SIO thin films on GSO, STO, and LSAT are coherently strained. As the compressive strain increases from GSO to LSAT, the 103 film reflection shifts downward, which is consistent with the $\theta$-$2\theta$ scans in Fig. 2a. The SIO thin film on MgO is relaxed due to the large lattice mismatch (~5%), and this sample is expected to emulate properties of unstrained ('bulk') SIO. The lattice parameters are found using the reciprocal space reflection peaks ($Q_i = 2\pi/d_{hkl}$, where $i \equiv \perp$ or $//$ and $d_{hkl}$ is the spacing between appropriate reflection planes (*hkl*)). Figure 2d shows the change of *in-plane* (*a*) and *out-of-plane* (*c*) lattice parameters as a function of the substrate in-plane lattice constant. The ratio *c/a* quantifies the tetragonal distortion due to strain present in the SIO thin films, with *c/a > 100%* corresponding to compressive strain. Figure 2e shows the thin films grown on LSAT, STO, and GSO are compressively strained with a tetragonal distortion, and the SIO thin film on MgO is relaxed with a cubic structure.

## III.    TRANSPORT PROPERTIES

Figure 3 shows temperature dependent resistivity curves $\rho(T)$, which indicate a large dependence on compressive lattice strain.  The resistivity of each strained SIO thin film systematically increases with in-plane compression.  The resistivity of the SIO films are of the same order of magnitude as the resistivity measured in a polycrystalline, high pressure synthesis study of SIO (~ 2 mΩ cm at room temperature).[12]  In addition to an increase in resistivity, the SIO films synthesized on STO and LSAT have an upturn in their resistivity curves at 175 K and 225 K, respectively.  The room temperature-normalized resistivity curves are shown in the bottom panel of Fig. 3.  The resistivity of the relaxed SIO on MgO has a positive slope (d$\rho$/d$T$) for all temperatures from 4 to 300 K, which is expected for metallic behavior.  The positive slope is reduced for the thin film grown on GSO and becomes increasingly negative for the thin films grown on STO and LSAT.  To gain a phenomenological understanding of the insulating regions, resistivity data (< 225 K) of SIO grown on STO and LSAT are fit with:

$$\rho = \rho_0 - \alpha T^{3/4} + \beta T^{3/2} \qquad (1)$$

where $\rho_0$, $\alpha$, and $\beta$ are the fit parameters for remnant resistance, three dimensional weak localization and inelastic scattering due to electron-boson interactions, respectively.[15, 16]  Weak localization is a disorder driven effect due to quantum interference of the conducting charge carriers at defect sites.[17]  The weak localization coefficient $\alpha$ for the SIO thin film on LSAT is twice as large as the fit coefficient for the thin film grown on STO, as shown in Table I.  This weak-localization model fit is similar to a previous study of compressively strained SIO perovskite films grown on LaAlO$_3$ substrates.[16]

## IV. OPTICAL PROPERTIES

Optical absorption spectroscopy provides insight into understanding the electronic band structure and free carrier dynamics of thin films. It also offers a method of determining the insulating gap energy by measuring the photon energy of zero absorption coefficient. Optical transmission spectra $T(\omega)$ of our samples are taken at room temperature using a Fourier-transform infrared spectrometer (for spectra regions between 50 meV – 0.6 eV) and a grating-type spectrophotometer (for spectra regions between 0.5 – 6 eV). The optical absorption spectra $\alpha(\omega)$ of SIO thin films are obtained using $\alpha(\omega) = -\frac{1}{t}\ln[T(\omega)_{film+substrate}/T(\omega)_{substrate}]$, where $t$ is the thin film thickness. According to Fermi's golden rule, $\alpha(\omega)$ is proportional to the product of the transition probability amplitude and joint density of states. Therefore, the $i^{th}$ interband-transition peak position $\omega_{0,i}$ in $\alpha(\omega)$ corresponds to the energy difference between bands, and the corresponding peak width $\gamma_i$ is proportional to the bandwidth. SIO thin films grown on MgO substrates are known to show a Drude-like response at energies in the infrared, an interband optical transition $\omega_{o,\beta} \sim 0.8\ eV$ from the $J_{eff} = 3/2$ band to the $J_{eff} = 1/2$ band, and a higher energy charge transfer from O $2p$ bands to Ir $5d$ bands.[1]

Despite the compressively strained thin films having a negative d$\rho$/d$T$ behavior, the absorption spectra in Fig. 4 show all the SIO films have a Drude-like response at low energies, which suggests a metallic behavior. The absorption spectra for all samples are fit to the Drude model for the low energy metallic response, a Lorentz oscillator for the interband β transition, and a second Lorentz oscillator for the higher energy charge transfer transition. As summarized in Table I, the plasma frequency $\omega_p$ is obtained from the Drude model as a fit parameter, and the free charge carrier density $n$ for each SIO film is found using $\omega_p = (n\ e^2/(m^*\varepsilon_0))^{1/2}$, where $e$ is the elementary charge, $m^*$ is the effective mass of an electron, and $\varepsilon_0$ is the permittivity of free space. S. J. Moon *et al.* reported a frequency-dependent mass enhancement of SIO's conducting carriers only at low energies, and the effective mass approaches the free electron mass $m_e$ around 0.08 eV.[1] Therefore, in our spectral range (> 0.2 eV) it is assumed $m^* = m_e$, and the free charge density $n$ is found to be $7.7 \times 10^{22}\ cm^{-3}$, $4.0 \times 10^{22}\ cm^{-3}$, and $4.9 \times 10^{21}\ cm^{-3}$ for SIO grown on MgO, STO, and LSAT, respectively. Although both compressively strained SIO films feature insulating properties, all films have free carrier concentrations of the same order of magnitude, albeit smaller than that of most metals. The carrier density of SIO on MgO is also verified using a room temperature Hall measurement and is found to be $7.3 \times 10^{22}\ cm^{-3}$, which is in agreement with the $n$ found in the optical spectra and supports the assumption $m^* = m_e$. The absorption peak $\omega_{o,\beta}$ around *0.8 eV* remains relatively constant for all SIO films; however, the peak width $\gamma_\beta$ of the relaxed SIO on MgO is larger than $\gamma_\beta$ of each compressively strained film on STO and LSAT. This is consistent with the expectation of in-plane compressive strain causing a decrease in the Ir-O-Ir bond angle and therefore decreasing the bandwidth. However, the reduction in bandwidth caused by compressive strain is not large enough to open an insulating band gap in this material.

## V. MAGNETO-RESISTANCE MEASUREMENTS

The SIO thin films grown under compressive strain (on STO and LSAT) exhibit a negative magneto-resistance (MR) behavior at low temperatures while the MR of the SIO thin film grown on MgO shows a positive $B^2$ behavior [MR (%) $= 100 \times (R(B) - R(0))/R(0)$, where $R(0)$ and $R(B)$ are thin film resistance for zero and non-zero magnetic field $B$, respectively]. For each sample, the MR response is qualitatively isotropic in the following field and thin film orientations: $B \parallel c, B \perp I$ (Fig. 5) and $B \perp c, B \perp I$, where $I$ is the in-plane current and $c$ is the out-of-plane film axis. The positive $B^2$ MR is typical for simple metals and is observed for the unstrained SIO on MgO. The negative MR for the compressively strained thin films grown on STO and LSAT is consistent with the data reported in Ref. [16]. The MR data supports the picture in Fig. 1 that weak localization causes the insulating behavior of compressively strained SIO thin films. For weak localization, time reversal symmetry must be preserved. By applying magnetic fields, a phase shift occurs in the electronic wave functions. Hence, this asymmetrical phase shift between forward and backscattered electrons suppresses the interference, i.e. weak localization, and results in a decrease in the resistance under magnetic fields, which is a negative MR behavior.[17]

## VI. SUMMARY

Epitaxial $SrIrO_3$ thin films have been synthesized on various substrates inducing in-plane compressive strain and film relaxation. Relaxed SIO is metallic for all temperatures while the strained thin films have an increase in their resistivity as a function of compression. Weak localization corrections are fit to the resistivity curves of the SIO thin films with insulating properties. Although the compressively strained films exhibit insulating properties, the optical absorption spectra of all SIO thin films have a Drude-like metallic response at low energies. The $J_{eff} = 3/2$ absorption peak is centered around 0.8 eV, and the bandwidth decreases with in-plane compressive strain. This implies that compressive strain induces an in-plane rotation of the oxygen octahedra causing a decrease in bandwidth, but fails to open an insulating bandgap. The positive, quadratic MR for the relaxed SIO film is typical for metals, and the negative MR of the compressively strained films is indicative for insulating samples due to weak localization. As schematically summarized by Fig. 1, these results demonstrate the effect of lattice strain on the electronic structure of SIO, and the role of weak localization in obtaining insulating states in strongly correlated, metallic systems. This offers an important progression in our understanding of similar *5d* transition metal oxide systems and heterostructures with strong spin-orbit coupling and electron correlation.

## ACKNOWLEDGMENTS

The authors would gratefully like to thank the National Science Foundation through Grant Nos. EPS-0814194 (the Center for Advanced Materials), DMR-1262261 (JWB), DMR-0856234 (GC), and DMR-1265162 (GC), and the Kentucky Science and Engineering Foundation with the Kentucky Science and Technology Corporation through Grant Agreement No. KSEF-148-502-12-303 (SSAS).

TABLE I. SIO thin film summary: tetragonal distortion (*c/a*) from reciprocal space maps; weak localization $\alpha/\rho_0$ and electron-boson scattering $\beta/\rho_0$ fit parameters from transport; and plasma frequency $\omega_p$, interband absorption $\omega_{o,\beta}$, and carrier density *n* from Drude-Lorentz fit of absorption spectra.

| Substrate | *c/a* (%) | Transport | | Absorption Spectra | | |
|---|---|---|---|---|---|---|
| | | $\alpha/\rho_0$ ($K^{-3/4}$) | $\beta/\rho_0$ ($10^{-5} K^{-3/2}$) | $\omega_p$ (eV) | $\omega_{o,\beta}$ (eV) | *n* ($10^{22}\ cm^{-3}$) |
| MgO | 100 | - | - | 10 | 0.91 | 7.7 |
| STO | 102 | 0.003 | 3.66 | 7.4 | 0.79 | 4.0 |
| LSAT | 103 | 0.006 | 5.33 | 2.6 | 0.79 | 0.49 |

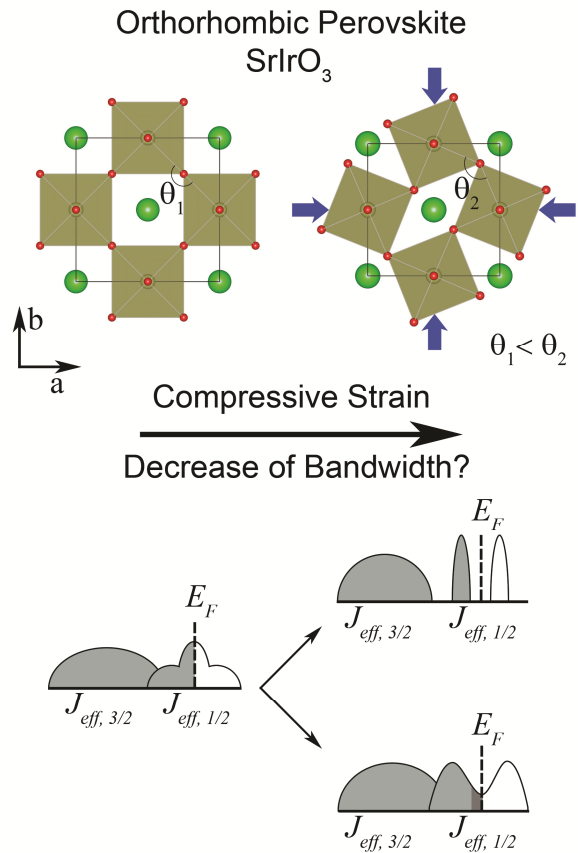

FIG. 1. Schematic diagrams showing the relationship between compressive strain and bandwidth reduction. (top panel) In-plane compressive strain will cause the Ir-O-Ir bond angle $\theta$ to decrease. (bottom panel) By reducing the bandwidth, the correlated metallic band structure of $SrIrO_3$ (left) can turn insulating either by (upper) a Mott gap opening or (lower) localized states (emphasized in dark gray) forming below the mobility edge. (color online)

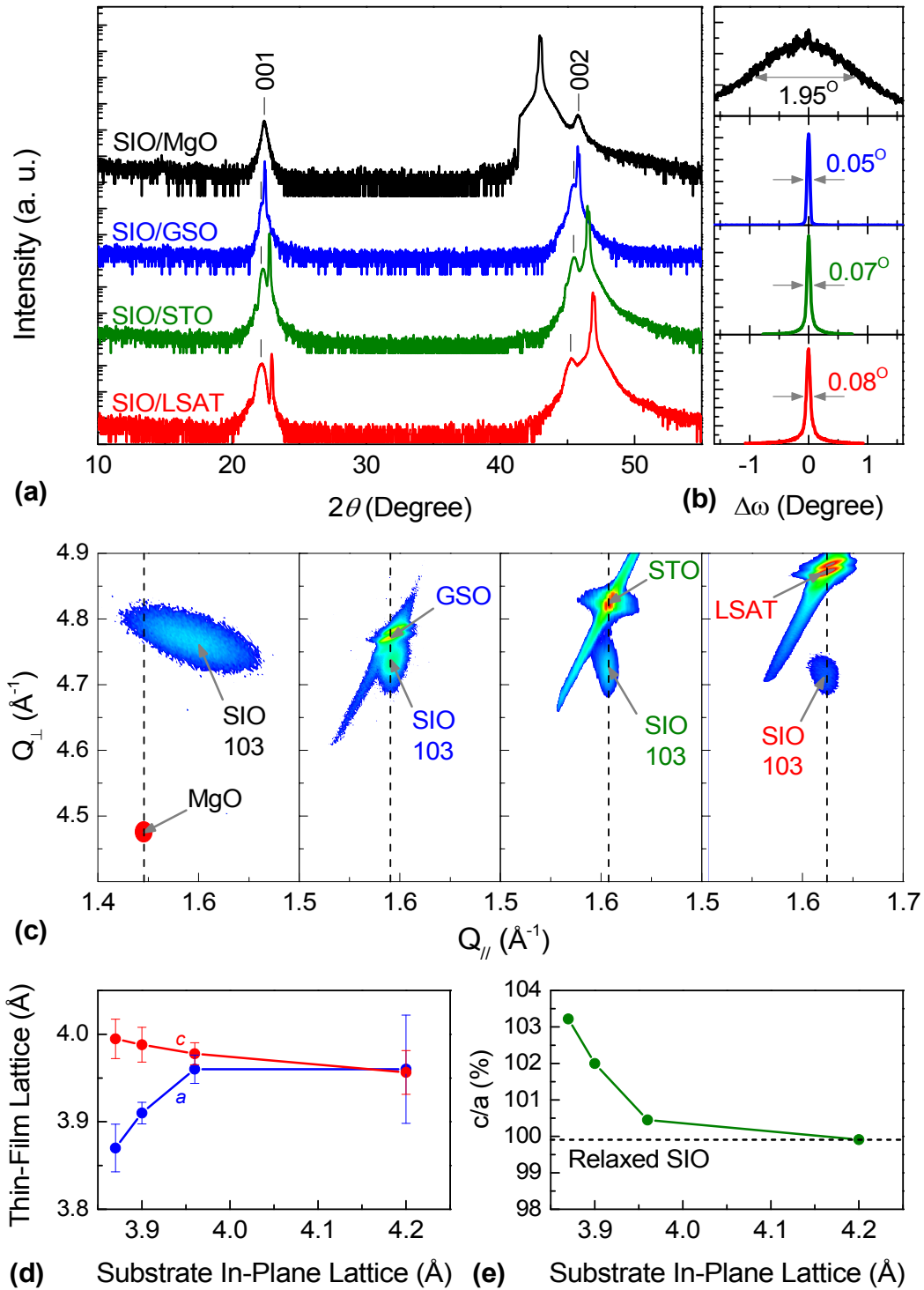

FIG. 2. X-ray diffraction scans confirming the epitaxial growth of perovskite SIO thin films on MgO (black), GSO (blue), STO (green), and LSAT(red): (a) $\theta$-$2\theta$ diffraction scans show correct $00l$ oriented film peaks, (b) rocking curve $\Delta\omega$ scans about the film's 002-reflection, and (c) reciprocal space maps around the 103-reflection (pseudo-cubic notation) of the substrates. (d) The out-of-plane $c$ and in-plane $a$ lattice constants of the strained SIO thin films and (e) the ratio $c/a$ as a function of substrate in-plane lattice parameters. (color online)

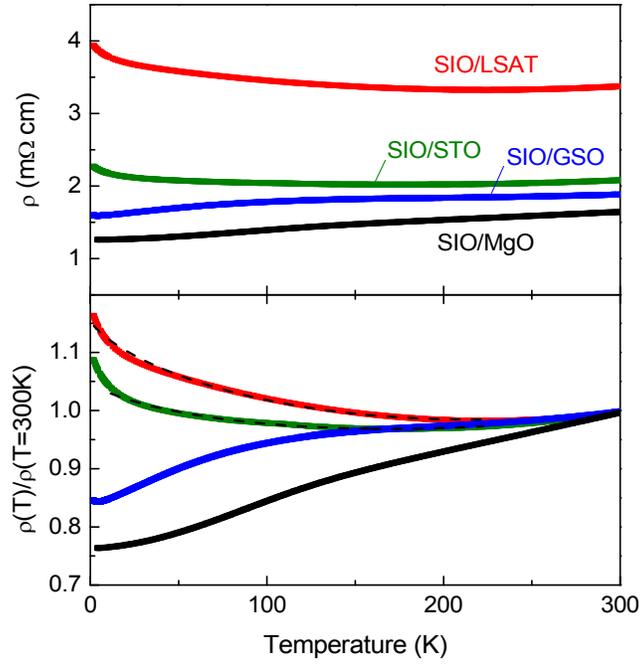

FIG. 3. Resistivity versus temperature (upper panel) of SIO films grown on MgO, GSO, STO, and LSAT substrates. The resistivity increases with in-plane compression. The room-temperature normalized resistivity (bottom panel) clearly shows the strain-dependence of the transport, and the insulating regions of SIO on STO and LSAT are fit (dashed lines) with a weak localization model. (color online)

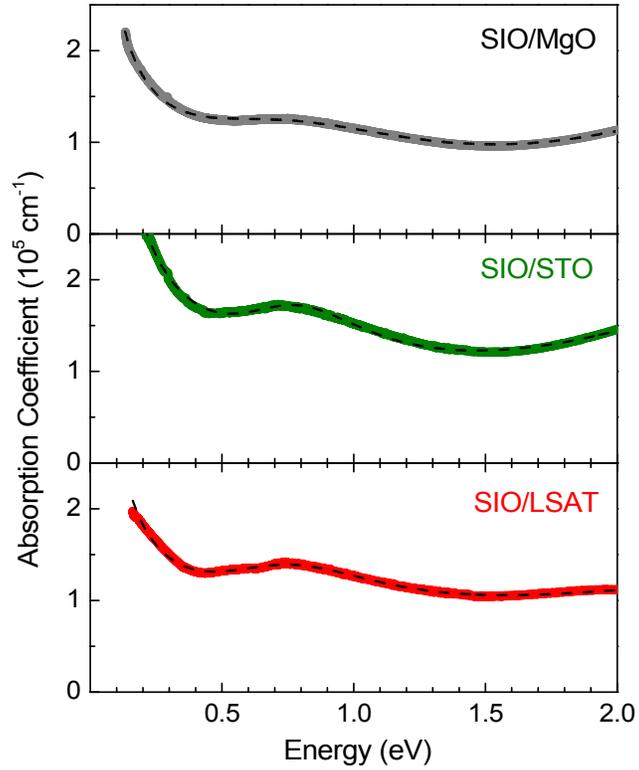

FIG. 4. Optical absorption spectra of SIO films grown on MgO, STO, and LSAT substrates. The dashed black lines are Drude-Lorentz fits using the Drude model and two Lorentz oscillators: for the interband $\beta$ absorption (~0.8 eV) and for the charge transfer. (color online)

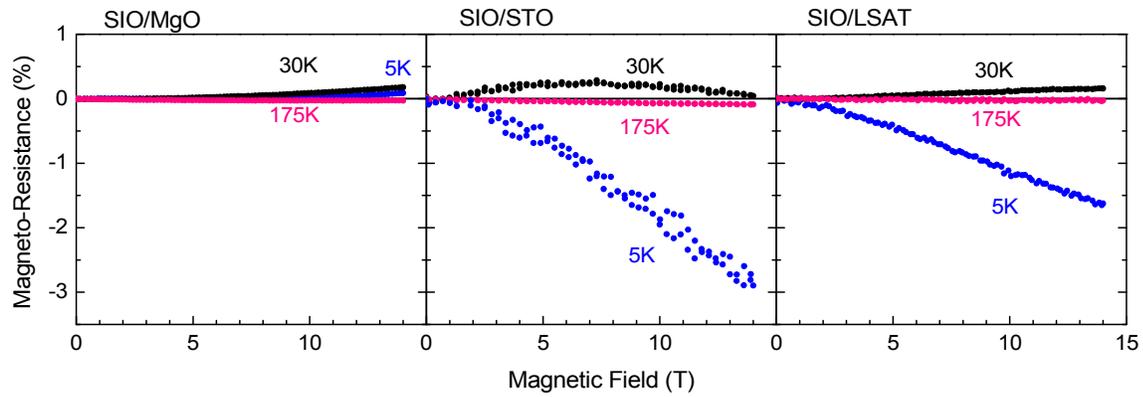

FIG. 5. Magneto-resistance (MR) versus magnetic field taken at 5K, 30K, and 175K for SIO films grown on MgO, STO, and LSAT. The SIO on MgO displays typical metallic $B^2$ behavior. The compressively strained films grown on STO and LSAT have negative MR, supporting weak localization. (color online)